# Investigation of a high-entropy oxide photocatalyst for hydrogen generation by first-principles calculations coupled with experiments: Significance of electronegativity

Jacqueline Hidalgo-Jiménez[a,b], Taner Akbay[c], Tatsumi Ishihara[a,d,e] and Kaveh Edalati[a,e,]*

[a] WPI, International Institute for Carbon-Neutral Energy Research (WPI-I2CNER), Kyushu University, Fukuoka, Japan
[b] Graduate School of Integrated Frontier Sciences, Department of Automotive Science, Kyushu University, Fukuoka, Japan
[c] Materials Science and Nanotechnology Engineering, Yeditepe University, Istanbul, Turkey
[d] Department of Applied Chemistry, Faculty of Engineering, Kyushu University, Fukuoka, Japan
[e] Mitsui Chemicals, Inc. - Carbon Neutral Research Center (MCI-CNRC), Kyushu University, Fukuoka, Japan

High-entropy oxides (HEOs), containing at least five principal cations, have recently emerged as promising photocatalysts for hydrogen production via water splitting. Despite their high potential, the impact of the cation mixtures on photocatalytic activity remains poorly understood. This study investigates the high-entropy photocatalyst $TiZrHfNbTaO_{11}$ using first-principles calculations combined with experimental methods to elucidate the effects of various elements on electronic structure and water splitting performance. The results indicate that the HEO exhibits a bandgap comparable to $TiO_2$ polymorphs rutile, brookite and anatase. Cations with lower electronegativity, such as hafnium and zirconium, provide the strongest water adsorption energy, serving as active sites for water adsorption. Additionally, the co-presence of highly electronegative cations like niobium and tantalum adjacent to hafnium and zirconium enhances charge transfer to water molecules, improving splitting efficiency. These findings suggest novel strategies for designing high-entropy photocatalysts by synergistic incorporating cations with different electronegativities.

***Keywords:*** photocatalysis; density functional theory (DFT); hydrogen production; high-entropy ceramics; high-pressure torsion (HPT)

*Corresponding author (E-mail: kaveh.edalati@kyudai.jp)

Global warming is becoming more evident day-to-day, and the utilization of fossil fuels is one of the main reasons for this environmental problem. Considering this reality, multiple alternatives have been developed to reduce the consumption of fossil fuels, one of them is the utilization of hydrogen as an energy carrier. However, hydrogen is mainly produced by fossil fuels which makes its production pathway environmentally harmful. Therefore, methods such as photocatalysis are receiving attention for producing hydrogen with a minor $CO_2$ footprint. Photocatalysis (Fig. 1(a)) utilizes sunlight to perform different reactions in the presence of a catalyst with the capability to create photogenerated excitons [1]. Hydrogen evolution from water splitting is one of the photocatalytic reactions and is still the main attraction of this research field.

Following a publication in 2020 [2]. High-entropy oxides (HEOs) have appeared as a new type of photocatalysts with high activity for various applications such as hydrogen evolution [2-4], oxygen production [5], $CO_2$ conversion [6-9] and dye degradation [10,11]. The HEOs are compounds containing five or more principal cations and oxygen anions [12,13]. Some other anions like nitrogen are sometimes added to HEOs to make more complex compounds such as high-entropy oxynitrides [12,13]. The large number of elements in HEOs enhances the entropy of mixing and leads to an entropy stabilization effect due to the reduction of Gibbs energy.

Despite the importance of entropy-induced stability in long-term photocatalysis, there is no clear relationship between the activity of photocatalysts and their entropy. It is believed that the existence of a large number of cations in HEOs produces complex mixtures of elements with different electronic structures and can lead to novel photocatalytic properties through the cocktail effect [2,6,14]. However, there have been few studies to clarify the electronic structure of these materials and compare their differences with conventional photocatalysts. Such studies are essential for the future design of high-entropy photocatalysts which are known to have high flexibility for tunning the composition and crystal structure for a specific application.

This study aims to clarify the characteristics of the band structure and active sites for a HEO using density functional theory (DFT) calculations. For the first study on this topic, a HEO TiZrHfNbTaO$_{11}$ was selected because earlier studies confirmed its activity for hydrogen production [2,15] and $CO_2$ conversion [6]. It is shown that the bandgap of this HEO and the adsorption energy of water on its surface are not much different from $TiO_2$, but the main difference is that the presence of hybridized orbitals enhances the charge transfer for photocatalysis [6]. This charge transfer feature together with the presence of multiple adsorption sites for water splitting indicates a high potential of HEOs not only for high photocatalytic activity but also for removal of co-catalyst addition, as experimentally reported in a recent publication [15].

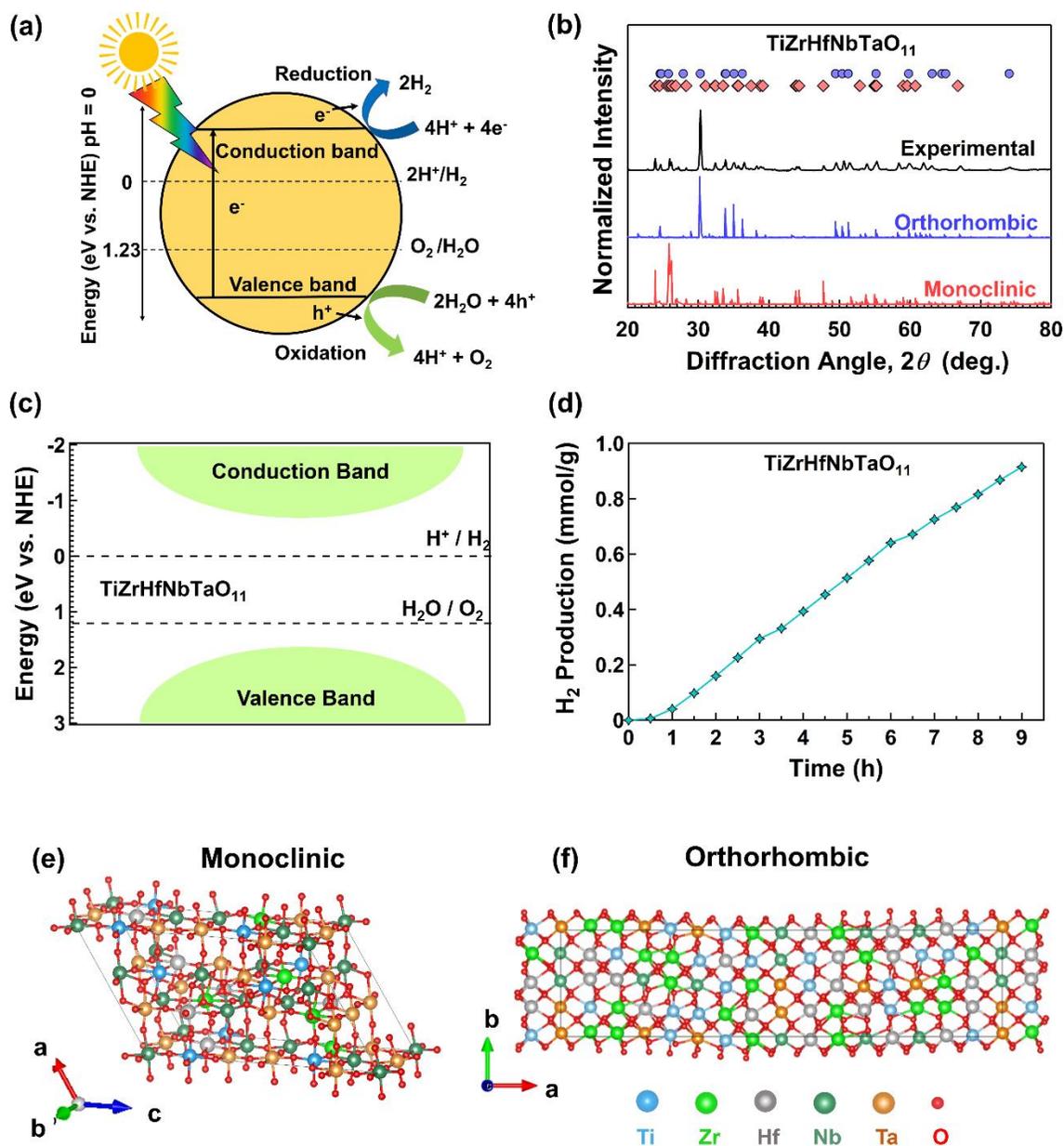

Fig. 1. (a) Schematic representation of photocatalytic water splitting. (b) Comparison of experimental XRD pattern of $TiZrHfNbTaO_{11}$ with calculated patterns for monoclinic and orthorhombic structures. (c) Experimental band structure of $TiZrHfNbTaO_{11}$ achieved by UV-Vis and UV photoelectron spectroscopy. (d) Hydrogen production from water splitting on $TiZrHfNbTaO_{11}$ photocatalyst. (e,f) Computed supercells of monoclinic and orthorhombic structures with compositions of $TiZrHfNb_3Ta_3O_{21}$ and $Ti_2Zr_2Hf_2NbTaO_{17}$, respectively.

To confirm the crystal structure, electronic band states and photocatalytic activity of $TiZrHfNbTaO_{11}$, the material was first experimentally synthesized and examined by different characterization methods. For the synthesis, titanium, zirconium, hafnium, niobium and titanium with high purity (99.5% for hafnium and >99.9% for other elements) were first mixed using the high-pressure torsion method (100 turns, 6 GPa pressure at room temperature) and then oxidized (24 h at 1373 K). High-pressure torsion was used in this study because of its high-strain effect on rapid mechanical alloying [16] and the influence of pressure-strain-defect interactions on accelerated phase transformations, as demonstrated experimentally and through multiscale atomistic and continuum theories and computational modeling [17,18]. One disadvantage of this method is the small surface area of the final powders, but attempts have been made to use laser fragmentation [19] or ball milling [15] to enhance the surface area. The crystal structure of the synthesized HEO was examined by X-ray diffraction (XRD) using the Cu Kα radiation, as shown in Fig. 1(b). XRD analysis shows good agreement with a previous study, showing that TiZrHfNbTaO$_{11}$ is a dual-phase oxide with the majority of the monoclinic structure and a minority of the orthorhombic structure [2]. An earlier study identified the monoclinic phase with a crystal structure of AB$_2$O$_7$ (space group: *A2/m*), while the orthorhombic phase has A$_6$B$_2$O$_{17}$ structure (space group: *Ima2*), substituting the A-site with titanium, zirconium and hafnium and the B-site with niobium and tantalum [2]. The powder diffraction patterns of both phases are consistent with the obtained XRD pattern (Fig. 1(b)). The band structure was examined by coupled ultraviolet-visible light (UV-vis) absorbance spectroscopy and UV photoelectron spectroscopy, as attempted in an earlier publication [2]. The band structure of the synthesized HEO, obtained by UV photoelectron spectroscopy and UV-vis spectroscopy, is shown in Fig. 1(c), indicating a bandgap of 3 eV with appropriate valence and conduction band levels for water splitting. The photocatalytic activity of the HEO for water splitting was examined using a xenon lamp with a power of 300W. For photocatalysis, 50 mg of TiZrHfNbTaO$_{11}$ was dispersed in a solution of 27 mL of deionized water and 3 mL of methanol and 0.0513 ml of Pt(NH$_3$)$_4$(NO$_3$)$_2$ (0.01 M), and the production of hydrogen and oxygen was measured using a gas chromatograph. No oxygen was detected, but as shown in Fig. 1(d), the hydrogen is produced continuously within an irradiation period of 9 h. The current experimental results confirm the crystal and band structures of TiZrHfNbTaO$_{11}$ as an active photocatalyst for hydrogen production.

For DFT calculations, both AB$_2$O$_7$-type monoclinic and A$_6$B$_2$O$_{17}$-type orthorhombic phases were simulated, but the main focus was on the monoclinic phase which is the major phase. As elements in high-entropy materials tend to be randomly distributed, special quasi-random structure (SQS) [20] configurations were generated using the code ICET [21]. The correlation functions of several nearest-neighbor clusters were tested to achieve a fully random distribution. In total, 25 supercells were generated by SQS. Those with powder diffraction patterns similar to the experimental XRD profile were optimized to select the ones with minimum energy. The comparison of XRD patterns of the selected SQS with the synthesized material is shown in Fig. 1(b), suggesting good agreement between the experiments and calculations. A comparison between the experimental and calculated lattice parameters of the two phases, as given in Table 1, also confirms good consistency despite the low symmetry of the phases. Fig. 1(e) and (f) show the supercells with 120 and 190 atoms visualized by using the software VESTA [22] for monoclinic and orthorhombic phases with compositions of TiZrHfNb$_3$Ta$_3$O$_{21}$ and Ti$_2$Zr$_2$Hf$_2$NbTaO$_{17}$, respectively. DFT calculations were performed using the Vienna Ab-initio Simulation Package (VASP) [23]. To describe exchange-correlation, generalized gradient approximation and Perdew-Burke-Ernzerhof (PBE) functional [24] were utilized. Hybrid functionals were not used due to the limitations in the computational power which increase each calculation time to months. Projected-augmented wave [25] of Ti_sv, Zr_sv, Hf_pv, Nb_sv, Ta_pv, O and H provided by VASP were utilized for the construction of supercells produced by SQS. The supercells were optimized before self-consistent field (SCF) and density of states (DOS) calculations with electronic and ionic constraints of $1.0 \times 10^{-6}$ eV and $1.0 \times 10^{-5}$ eV, respectively. To examine the interaction of water with the different cations on the surface of catalysts, a Python code was implemented to generate a slab of the (100) atomic plane of the monoclinic structure, as the major phase (Fig. 2(a)). This termination was selected due to its predominant intensity in the calculated XRD pattern, although finding the atomic planes with the highest photocatalytic activity remains a computationally expensive topic for future studies. The slab model consisted of three layers of atoms where the center layer was fixed to include a contribution from the bulk. The total ionic and Hartree potential of the slab in 20 Å vacuum was analyzed to avoid polarizability and to minimize the unintended interaction in the pristine model due to the periodicity, as shown in Fig. 2(b). Subsequently, one water molecule was located over the different cations at a distance of 2.5 Å to estimate the binding energy between water and the catalyst surface. The final model was optimized allowing the movement of atoms in the bottom and top layers of the slab and the water molecule. The electronic and ionic constraints and other parameters utilized for the optimization were the same as those utilized for optimizing the supercells.

Table 1. Comparison of calculated and experimental lattice parameters for monoclinic and orthorhombic phases of $TiZrHfNbTaO_{11}$.

|  | Monoclinic | | Orthorhombic | |
| --- | --- | --- | --- | --- |
|  | Experimental | Calculated | Experimental | Calculated |
| *a* (Å) | 11.93 | 12.04 | 40.92 | 41.02 |
| *b* (Å) | 3.81 | 3.83 | 4.93 | 4.95 |
| *c* (Å) | 20.44 | 20.59 | 5.27 | 5.29 |
| *α* (°) | 90 | 89.89 | 90 | 90.00 |
| *β* (°) | 120.16 | 119.85 | 90 | 90.00 |
| *γ* (°) | 90 | 90.06 | 90 | 90.00 |

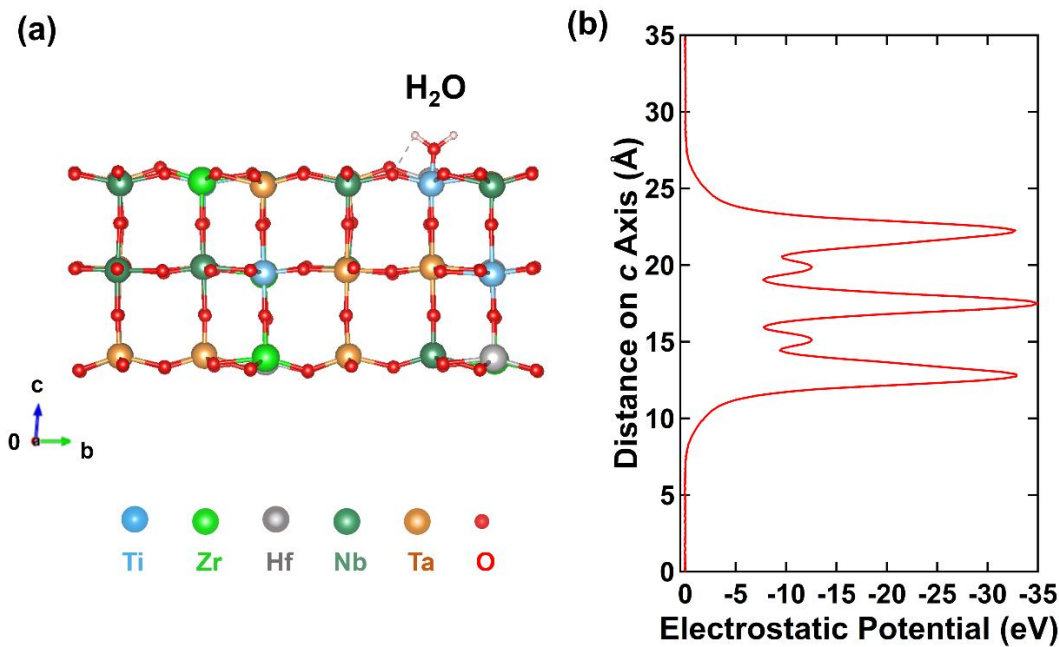

Fig. 2. (a) Surface model of the monoclinic phase with $TiZrHfNb_3Ta_3O_{21}$ composition for the atomic plane (100) with a water molecule on top. (b) Total potential variation along the normal direction of the surface model without surface water molecule.

The DOS plots for the two phases of the HEO are shown in Fig. 3 which indicates three important points: (i) In both phases the conduction band is mostly formed by cations. In the monoclinic phase, almost all cations have a similar influence on the valence band, while in the case of orthorhombic, titanium has a major contribution to the valence band. (ii) Both phases show asymmetry between the spin-up and spin-down DOS results for niobium atoms, suggesting the presence of dipole moments [26]. The dipole moments were shown to be beneficial for water-splitting reactions [27]. Similar behavior was reported for niobium in other oxides [28]. (iii) The bandgap for the monoclinic and orthorhombic phases are 2.3 eV and 1.7 eV, respectively, which are comparatively low compared to the experimental value of 3.0 eV. The low value obtained for the bandgap in DFT is related to the underestimation of Coulomb interactions [29,30]. Table 2 shows the comparison of the bandgap obtained for the monoclinic and orthorhombic structures of HEO with $TiO_2$ polymorphs of anatase, rutile, columbite and brookite [31,32]. $TiO_2$ polymorphs also show a smaller calculated bandgap compared to the experimental values. However, a comparison between the bandgap of the monoclinic phase of HEO with the ones calculated for $TiO_2$, confirms that HEO cannot be considered a low-bandgap photocatalyst which is consistent with earlier experimental reports [2,15].

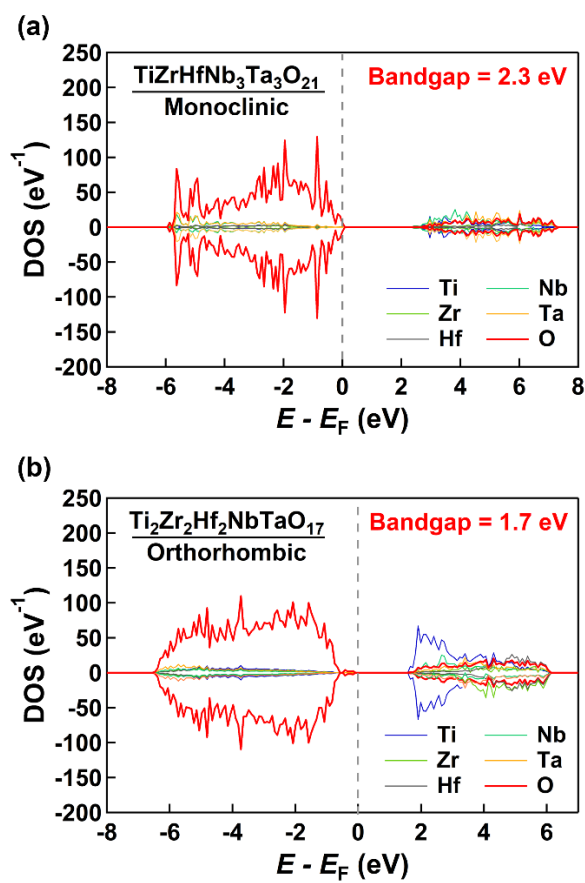

Fig. 3. Calculated DOS plots including the contribution of each element in the band structure of high-entropy oxide $TiZrHfNbTaO_{11}$ for (a) monoclinic and (b) orthorhombic phases.

Table 2. Comparison of calculated and experimental electronic bandgap of TiZrHfNbTaO$_{11}$ with TiO$_2$ polymorphs anatase, rutile, columbite and brookite.

| | Bandgap (eV) | | | | |
|---|---|---|---|---|---|
| | TiZrHfNbTaO$_{11}$ | Anatase | Rutile | Columbite | Brookite |
| DFT | 2.3 (Monoclinic), 1.7 (Orthorhombic) | 2.3 | 1.7 | 2.6 | 2.4 |
| Experimental | 3.0 | 3.2 | 3.0 | --- | 3.4 |
| Reference | This study | [1,31] | [31] | [31] | [31,32] |

The presence of at least five cations in high-entropy photocatalysts leads to different atomic-scale chemical environments and, thus, to different interactions with reactant molecules on the surface. In an attempt to show the effect of the chemical environment on the absorption of the water molecule on the TiZrHfNbTaO$_{11}$ surface, the charge distribution visualization on different cations was conducted for the monoclinic phase as shown in Fig. 4. The blue isosurface in this figure correspond to the electron depletion and the yellow isosurface correspond to electron accumulation regions. Fig. 4 indicates that the charge transfer and the adsorption of water to the surface depend on three chemical environment parameters: (i) the cation on which a water molecule is placed, (ii) surface and subsurface oxygen atoms close to the water molecule, and (iii) neighboring surface cations to the cation that hosts the water molecule. These three parameters are discussed below:

- Charge transfer strongly depends on the cation beneath the water molecule, and this subsequently affects the water adsorption on the surface. The interaction of the water molecule and the surface can be quantified through the following equation by comparing the energy of the water molecule and surface before adsorption ($E_{H2O}$ and $E_{surface}$) and after adsorption ($E_{surface+H2O}$).

$$E_{ads} = E_{surface+H2O} - (E_{surface} + E_{H2O}) \qquad (1)$$

Table 3 shows the water adsorption energy ($E_{ads}$) calculated using equation (1) by placing the water molecule on different cations in Fig. 4. Hafnium shows the strongest adsorption energy which suggests that it can act as the most active site for the adsorption of water. The binding energy of water to zirconium is also strong, but other cations show somehow low adsorption energies. Table 3 shows that there is an inverse relation between the electronegativity of cations (taken from [33]) and water adsorption energy: lower electronegativity leads to higher adsorption energy. Overall electron distribution analysis shown in Fig. 5 confirms that the charge distribution and transfer between the surface of the catalyst and water molecule becomes more intense when water is placed on cations with low electronegativity such as hafnium. Comparatively, the distribution of electrons between niobium and water is weaker due to the high electronegativity of niobium.

- When a water molecule is placed on different cations, surface oxygen atoms show higher charge accumulation (bigger yellow isosurface in Fig. 4) than sub-surface oxygen atoms. This indicates that surface oxygen atoms contribute more significantly to water adsorption, while sub-surface oxygen atoms affect charge transfer. As will be discussed below, the contribution of oxygen to charge transfer to the water molecule depends on its surrounding cations.

- When the cation that is beneath the water molecule is next to atoms like niobium and tantalum as good electron donors with high electronegativity [28], a higher electron transfer is visible to water through surface oxygen. For example, in Fig. 4(a), (c) and (e), there is an electron cloud from niobium to the host cation and water molecule through surface oxygen, while the electron cloud from other cations is weaker. As a result, the water molecule rotates and one of its hydrogen atoms points towards niobium or tantalum. The transfer of the charges from niobium or tantalum to this hydrogen atom suggests that it has a higher tendency to split from water compared to the opposite hydrogen atom. Therefore, although water adsorption energy calculations in Table 3 suggest that niobium is not as active as other cations, its role in improving the charge transfer and splitting water is significant in TiZrHfNbTaO$_{11}$. The heterogeneity in charge transfer from different cations to the water molecule, which is beneficial for photocatalysis, can hardly be observed in binary oxide photocatalysts such as TiO$_2$ without the presence of defects and co-catalysts [34]. Such heterogeneities can be beneficial for avoiding the addition of a co-catalyst for water splitting [15].

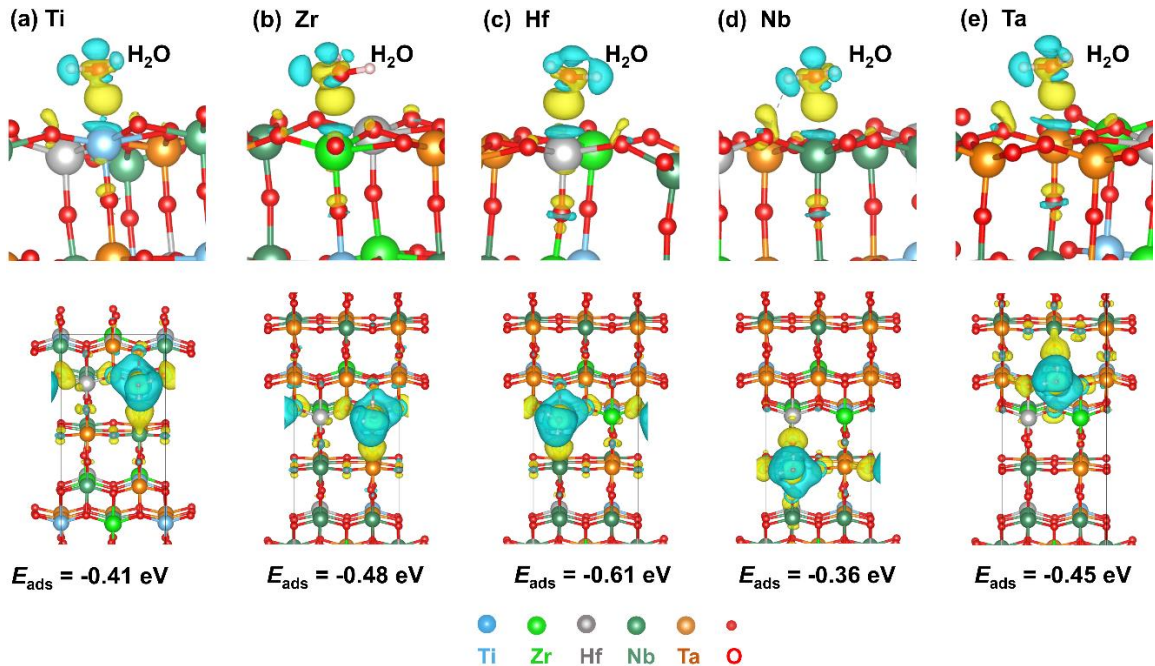

Fig. 4. Effect of chemical environment on localized charge distribution for water adsorption on (100) atomic plane of high-entropy oxide with monoclinic structure when a water molecule is located over (a) titanium, (b) zirconium, (c) hafnium, (d) niobium and (e) tantalum. Blue color corresponds to electron depletion, yellow color corresponds to electron accumulation, and upper and lower images show the cross-sectional and plan views, respectively. $E_{ads}$ refers to water adsorption energy on the selected cation. Isosurface color scales are -0.002 electrons per Å$^3$ for blue and +0.002 electrons per Å$^3$ for yellow.

Table 3. Adsorption energy ($E_{ads}$) of H$_2$O molecule placed on (100) atomic plane of high-entropy oxide with monoclinic structure over different cations with different electronegativity levels.

| Cation | Ti | Zr | Hf | Nb | Ta |
|---|---|---|---|---|---|
| **Electronegativity, $X$** | 1.54 | 1.33 | 1.30 | 1.60 | 1.50 |
| $E_{ads}$ (eV) | -0.41 | -0.48 | -0.61 | -0.36 | -0.45 |

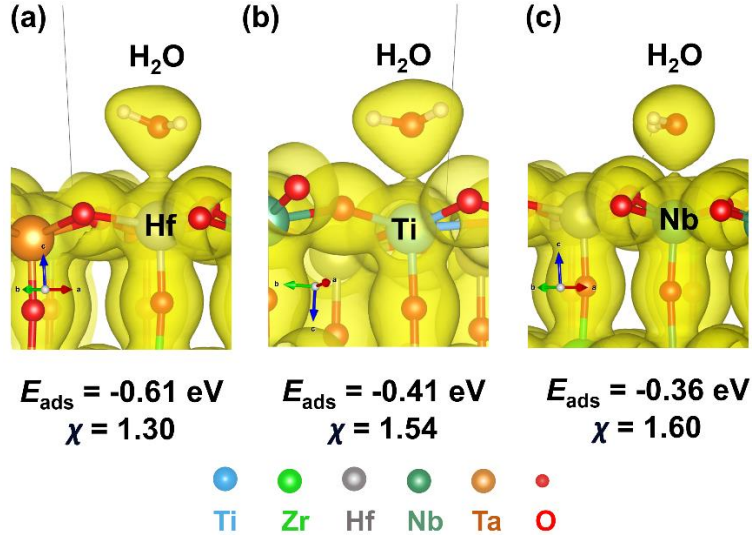

Fig.5. Overall electron distribution between the surface of high-entropy oxide with monoclinic structure and water molecule placed on (a) hafnium with strong interaction, (b) titanium with moderate interaction and (c) niobium with weak interaction. $E_{ads}$ refers to water adsorption energy on the selected cation and $X$ is electronegativity. Isosurface color scales are 0.033 electrons per Å$^3$.

The DFT calculations in this study suggest the importance of individual cations on the absorbance of water on the HEO surface. Here, a comparison of water adsorption energy on HEO with that on $TiO_2$ polymorphs as the most active photocatalysts for water splitting is beneficial. The adsorption energy calculated in this study for the anatase (101) atomic plane is -0.93 eV, while some other reported data in the literature for different $TiO_2$ polymorphs are given in Table 4 [35-41]. The comparison demonstrates that the adsorption energy of water over the titanium cation of the HEO is somehow smaller than in the $TiO_2$ anatase (101) plane. However, the adsorption energies are somehow higher than some reported data such as the anatase (100) plane [35] or rutile (110) plane [36]. This indicates that in addition to electronegativity, crystal structure and atomic configuration on crystallographic planes can affect the charge distribution of the surface and accordingly the adsorption energy. It should be noted the water adsorption energy of the most active cation of the HEO, i.e. hafnium, is still comparable to various reported data for $TiO_2$. In addition to $TiO_2$ polymorphs, the water adsorption energies reported for $ZrO_2$ [42] $HfO_2$ [43] and $Nb_2O_5$ [44] are also included in Table 4 (no data for $Ta_2O_5$ could be found). Although the data reported in different publications should be compared with care, there is an apparent trend of decreasing water adsorption energy with electronegativity which is consistent with this study.

The DFT calculations also suggest the importance of neighboring cations and oxygen atoms (on both surface and sub-surface) on heterogeneous charge transfer for the splitting of water molecules on high-entropy photocatalysts. For $TiO_2$, titanium gets reduced with the interaction with UV light from $Ti^{4+}$ to $Ti^{3+}$ making significant charge transfer to the surface of $TiO_2$ for photocatalysis [39,41,45]. Although $ZrO_2$, $Nb_2O_5$ and $Ta_2O_5$ are not as active as $TiO_2$ photocatalysts and $HfO_2$ is rarely considered a photocatalyst [46], it was shown that their addition to $TiO_2$ or other oxides as a dopant is effective for improving photocatalysis [47-50]. For example, it was shown that the doping of $TiO_2$ with tantalum or niobium provides extra electrons that are distributed among the other atoms modifying the reaction and charge separation behavior [47,48]. Hafnium as a dopant was reported to improve photocatalytic activity due to its charge separation and transfer effects and organic molecule adsorption enhancement [49]. The doping of $TiO_2$ with zirconium was also reported to enhance the charge transport and reduce the recombination of charge carriers [50]. Therefore, the co-existence of cations and their effect on the structure can positively affect the charge transfer and separation for water splitting, a fact that can be tuned by using HEOs. This indicates that a combination of transition metals, elements from the left side of the periodic table (i.e. alkaline metals with low electronegativity) and elements from the right side of the periodic table (e.g. bismuth and gallium with high electronegativity) can be a future strategy to create active high-entropy photocatalyst.

Table 4. Water adsorption energy comparison between high-entropy oxide with monoclinic structure and $TiO_2$ photocatalysts.

| Phase | $E_{ads}$ (eV) | Method | Software | Ref. |
|---|---|---|---|---|
| Anatase (101) | -0.78 | GGA-PBE | QE | [37] |
| Anatase (112) | -1.22 | GGA-PBE | QE | [37] |
| Anatase (100) | -0.65 | GGA-PBE | QE | [37] |
| Anatase (101) | -0.56 | GGA-PW91 | VASP | [35] |
| Anatase (100) | -0.24 | GGA-PW91 | VASP | [35] |
| Anatase (001) | -0.65 | GGA-PW91 | VASP | [38] |
| Anatase (101) | -0.69 | GGA-PBEsol | CASTEP | [39] |
| Rutile (110) | -0.23 | GGA-PBEsol | CASTEP | [36] |
| Rutile (100) | -1.33 | GGA-PBEsol | CASTEP | [36] |
| Rutile (001) | -1.42 | GGA-PBEsol | CASTEP | [36] |
| Anatase (110) | -0.69 | GGA-PBEsol | VASP | [40] |
| Anatase (101) | -1.30 | GGA-PBEsol | VASP | [41] |
| Rutile (110) | -1.50 | GGA-PBEsol | VASP | [41] |
| $ZrO_2$ (-111) | -1.20 to -0.83 | GGA-PBE | VASP | [42] |
| $ZrO_2$ (-101) | -1.50 to -1.21 | GGA-PBE | VASP | [42] |
| $HfO_2$ (110) | -1.79 | GGA-PW91 | $Dmol^3$ | [43] |
| $Nb_2O_5$ (010) | -1.10 | GGA-PBE | QE | [44] |
| Anatase (101) | -0.93 | GGA-PBE | VASP | This study |
| HEA (100) on Ti | -0.41 | GGA-PBE | VASP | This study |
| HEA (100) on Hf | -0.61 | GGA-PBE | VASP | This study |

In conclusion, the random distribution of the elements in high-entropy materials, such as $TiZrHfNbTaO_{11}$, as a new type of photocatalyst for water splitting provides different active sites with different water adsorption energies. This adsorption energy for a given atomic plane is inversely proportional to the electronegativity of the cation that makes bonding with water. Moreover, the presence of different cations on the surface provides a heterogeneous charge distribution and transfer which is beneficial for water splitting. Although the water adsorption energy for $TiZrHfNbTaO_{11}$ is somehow lower than $TiO_2$ polymorphs, the results in this study are of significance for the design of active high-entropy photocatalysts by considering the electronegativity and interaction of cations on the surface.

**Acknowledgments**

The author JHJ acknowledges a scholarship from the Q-Energy Innovator Fellowship of Kyushu University. This study is supported partly by Mitsui Chemicals, Inc., Japan, partly through Grants-in-Aid from the Japan Society for the Promotion of Science (JP19H05176, JP21H00150 &

**References**
[1]  A. Fujishima, K. Honda, Electrochemical photolysis of water at a semiconductor electrode, Nature 238 (1972) 37-38.
[2]  P. Edalati, Q. Wang, H. Razavi-Khosroshahi, M. Fuji, T. Ishihara, K. Edalati, Photocatalytic hydrogen evolution on a high-entropy oxide, J. Mater. Chem. A 8 (2020) 3814-3821.
[3]  P. Edalati, X.F. Shen, M. Watanabe, T. Ishihara, M. Arita, M. Fuji, K. Edalati, High-entropy oxynitride as a low-bandgap and stable photocatalyst for hydrogen production, J. Mater. Chem. A. 9 (2021) 15076-15086.
[4]  S. Nundy, D. Tatar, J. Kojcinovic, H. Ullah, A. Ghosh, T.K. Mallick, R. Meinusch, B.M. Smarsly, A.A. Tahir, I. Djerdj, Bandgap engineering in novel fluorite-type rare earth high-Entropy Oxides (RE-HEOs) with computational and experimental validation for photocatalytic water splitting applications, Adv. Sustain. Syst. 6 (2022) 1-20.
[5]  K. Edalati, A. Bachmaier, V.A. Beloshenko, Y. Beygelzimer, V.D. Blank, W. Botta, K. Bryla, J. Cizek, S. Divinski, N.A. Enikeev, Y. Estrin, G. Faraji, R.B. Figueiredo, M. Fuji, T. Furuta, T. Grosdidier, J. Gubicza, A. Hohenwarter, Z. Horita, J. Huot, Y. Ikoma, M. Janecek, M. Kawasaki, P. Kral, S. Kuramoto, T.G. Langdon, D. Leiva, V.I. Levitas, A. Mazilkin, M. Mito, H. Miyamoto, T. Nishizaki, R. Pippan, V.V. Popov, E.N. Popova, G. Purcek, O. Renk, A. Revesz, X. Sauvage, V. Sklenicka, W. Skrotzki, B.B. Straumal, S. Suwas, L.S. Toth, N. Tsuji, R.Z. Valiev, G. Wilde, M.J. Zehetbauer, X. Zhu, Nanomaterials by severe plastic deformation: review of historical developments and recent advances, Mater. Res. Lett. 10 (2022) 163-256.
[6]  S. Akrami, Y. Murakami, M. Watanabe, T. Ishihara, M. Arita, M. Fuji, K. Edalati, Defective high-entropy oxide photocatalyst with high activity for $CO_2$ conversion, Appl. Catal. B 303 (2022) 120896.
[7]  S. Akrami, P. Edalati, Y. Shundo, M. Watanabe, T. Ishihara, Significant $CO_2$ photoreduction on a high-entropy oxynitride, Chem. Eng. J. 449 (2022) 137800.
[8]  A. Haeussler, S. Abanades, J. Jouannaux, A. Julbe, Demonstration of a ceria membrane solar reactor promoted by dual perovskite coatings for continuous and isothermal redox splitting of $CO_2$ and $H_2O$, J. Memb. Sci. 634 (2021) 119387.
[9]  Y. Li, X. Bai, D. Yuan, C. Yu, X. San, Y. Guo, L. Zhang, J. Ye, Cu-based high-entropy two-dimensional oxide as stable and active photothermal catalyst, Nat. Commun. 14 (2023) 3171.
[10] M. Anandkumar, A. Lathe, A.M. Palve, A.S. Deshpande, Single-phase $Gd_{0.2}La_{0.2}Ce_{0.2}Hf_{0.2}Zr_{0.2}O_2$ and $Gd_{0.2}La_{0.2}Y_{0.2}Hf_{0.2}Zr_{0.2}O_2$ nanoparticles as efficient photocatalysts for the reduction of Cr (VI) and degradation of methylene blue dye, J. Alloys Compd. 850 (2021) 156716.
[11] F. Zhang, W. Zhou, Y. Zhang, Y. Lei, L. Wu, T. Liu, R. Fan, Spectroscopic analyses and photocatalytic properties of transition group metal oxide films with different entropy values, Mater. Sci. Semicond. Process. 169 (2024) 107928.
[12] S. Akrami, P. Edalati, M. Fuji, K. Edalati, High-entropy ceramics: review of principles, production and applications, Mater. Sci. Eng. R 146 (2021) 100644.
[13] C. Oses, C. Toher, S. Curtarolo, High-entropy ceramics, Nat. Rev. Mater. 5 (2020) 295-309.

[14] D. Berardan, A.K. Meena, S. Franger, C. Herrero, N. Dragoe, Controlled Jahn-Teller distortion in (MgCoNiCuZn)O-based high entropy oxides, J. Alloys Compd. 704 (2017) 693-700.

[15] O. Guler, M. Boyrazli, M.G. Albayrak, S.H. Guler, T. Ishihara, K. Edalati, Photocatalytic hydrogen evolution of TiZrNbHfTaO$_x$ high-entropy oxide synthesized by mechano-thermal method, Materials 17 (2024) 853.

[16] K. Edalati, Metallurgical alchemy by ultra-severe plastic deformation via high-pressure torsion process, Mater. Trans. 60 (2019) 1221-1229.

[17] V.I. Levitas, High-pressure phase transformations under severe plastic deformation by torsion in rotational anvils, Mater. Trans. 60 (2019) 1294-1301.

[18] V.I. Levitas, Recent in situ experimental and theoretical advances in severe plastic deformations, strain-induced phase transformations, and microstructure evolution under high pressure, Mater. Trans. 64 (2023) 1866-1878.

[19] Z.P. Tehrani, T. Fromme, S. Reichenberger, B. Gökcee, T. Ishihara, T. Lippert, K. Edalati, Laser fragmentation of high-entropy oxide for enhanced photocatalytic carbon dioxide ($CO_2$) conversion and hydrogen ($H_2$) production, Advanced Powder Technol. 35 (2024) 104448.

[20] A. Zunger, S.H. Wei, L.G. Ferreira, J.E. Bernard, Special quasirandom structures, Phys. Rev. Lett. 65 (1990) 353-356.

[21] M. Angqvist, W.A. Munoz, J.M. Rahm, E. Fransson, C. Durniak, P. Rozyczko, T.H. Rod, P. Erhart, ICET - a Python library for constructing and sampling alloy cluster expansions, Adv. Theory Simul. 2 (2019) 190015.

[22] K. Momma, F. Izumi, VESTA 3 for three-dimensional visualization of crystal, volumetric and morphology data, J. Appl. Crystallogr. 44 (2011) 1272-1276.

[23] G. Kresse, J. Furthmuller, Efficient iterative schemes for ab initio total-energy calculations using a plane-wave basis set, Phys. Rev. B 54 (1996) 11169-11186.

[24] J.P. Perdew, K. Burke, M. Ernzerhof, Generalized gradient approximation made simple, Phys. Rev. Lett. 77 (1996) 3865-3868.

[25] P.E. Blochl, Projector augmented-wave method, Phys. Rev. B, 50 (1994) 17953-17979.

[26] Y. Inoue, Photocatalytic water splitting by $RuO_2$-loaded metal oxides and nitrides with $d^0$- and $d^{10}$-related electronic configurations, Energy Environ. Sci. 2 (2009) 364-386.

[27] Y. Liu, S. Yin, P.K. Shen, Asymmetric 3d electronic structure for enhanced oxygen evolution catalysis, ACS Appl. Mater. Interfaces 10 (2018) 23131-23139.

[28] Y. Liang, B. Luo, H. Dong, D. Wang, Electronic structure and transport properties of sol-gel-derived high-entropy Ba($Zr_{0.2}Sn_{0.2}Ti_{0.2}Hf_{0.2}Nb_{0.2}$)$O_3$ thin films, Ceram. Int. 47 (2021) 20196-20200.

[29] J. Hubbard, Electron correlations in narrow energy bands, Proc. R. Soc. London. Ser. A Math. Phys. Sci. 276 (1963) 238-257.

[30] M. Shishkin, H. Sato, DFT+ U in Dudarev's formulation with corrected interactions between the electrons with opposite spins: the form of Hamiltonian, calculation of forces, and bandgap adjustments, J. Chem. Phys. 151 (2019), 024102.

[31] J. Hidalgo-Jimenez, T. Akbay, T. Ishihara, K. Edalati, Understanding high photocatalytic activity of the $TiO_2$ high-pressure columbite phase by experiments and first-principles calculations, J. Mater. Chem. A 11 (2023) 23523-23535.

[32] J. Zhang, P. Zhou, J. Liu, J. Yu, New understanding of the difference of photocatalytic activity among anatase, rutile and brookite $TiO_2$, Phys. Chem. Chem. Phys. 16 (2014) 20382-20386.

[33] K.Y. Li, D.F. Xue,. New development of concept of electronegativity, Chinese Sci. Bull. 54 (2009) 328-334.

[34] B. Liu, X. Zhao, C. Terashima, A. Fujishima, K. Nakata, Thermodynamic and kinetic analysis of heterogeneous photocatalysis for semiconductor systems, Phys. Chem. Chem. Phys. 16 (2014), 8751-8760.

[35] A.S. Barnard, P. Zapol, L.A. Curtiss, Modeling the morphology and phase stability of $TiO_2$ nanocrystals in water, J. Chem. Theory Comput. 1 (2005) 107-116.

[36] Z.Y. Zhao, Single water molecule adsorption and decomposition on the low-index stoichiometric rutile $TiO_2$ surfaces, J. Phys. Chem. C 118 (2014) 4287-4295.

[37] X.Q. Gong, A. Selloni, Role of steps in the reactivity of the anatase $TiO_2$(101) surface, J. Catal. 249 (2007) 134-139.

[38] R. Erdogan, O. Ozbek, I. Onal, A periodic DFT study of water and ammonia adsorption on anatase $TiO_2$ (001) slab, Surf. Sci. 604 (2010) 1029-1033.

[39] S. Jia, J. Gao, Q. Shen, J. Xue, Z. Zhang, X. Liu, H Jia, Effect of oxygen vacancy concentration on the photocatalytic hydrogen evolution performance of anatase $TiO_2$: DFT and experimental studies, J. Mater. Sci. Mater. Electron. 32 (2021) 13369-13381.

[40] Q. Guo, C. Xu, Z. Ren, W. Yang, Z. Ma, D. Dai, H. Fan, T.K. Minton, X. Yang, Stepwise photocatalytic dissociation of methanol and water on $TiO_2$(110), J. Am. Chem. Soc. 134 (2012) 13366-13373.

[41] N. Hamamoto, T. Tatsumi, M. Takao, T. Toyao, Y. Hinuma, K.I. Shimizu, T. Kamachi, Effect of oxygen vacancies on adsorption of small molecules on anatase and rutile $TiO_2$ surfaces: a frontier orbital approach, J. Phys. Chem. C 125 (2021) 3827-3844.

[42] S.T. Korhonen, M. Calatayud, A.O. I. Krause, Stability of hydroxylated (-111) and (-101) surfaces of monoclinic zirconia: a combined study by DFT and infrared spectroscopy, J. Phys. Chem. C 122 (2008) 6469-6476.

[43] L. Li, X. Huang, Y.F. Zhang, X. Guo, W.K. Chen, First-principles investigation of $H_2O$ on $HfO_2$ (1 1 0) surface, Appl. Surf. Sci, 264 (2013) 424-432.

[44] M.B. Pinto, A.L. Soares, M.C. Quintão, H.A. Duarte, H. A. De Abreu, Unveiling the structural and electronic properties of the B-Nb2O5 surfaces and their interaction with $H_2O$ and $H_2O_2$, J. Phys. Chem. C 122 (2018) 6618-6628.

[45] K. Nakata, A. Fujishima, $TiO_2$ photocatalysis: design and applications, J. Photochem. Photobiol. C 3 (2012) 169-189.

[46] L.A. Gonzalez, S. Galvez-Barboza, E. Vento-Lujano, J.L. Rodriguez-Galicia, L.A. Garcia-Cerda, Mn-modified $HfO_2$ nanoparticles with enhanced photocatalytic activity, Ceram. Int. 46 (2020) 13466-13473.

[47] N. Kim, E.M. Turner, Y. Kim, S. Ida, H. Hagiwara, T. Ishihara, E. Ertekin, Two-dimensional $TiO_2$ nanosheets for photo and electro-chemical oxidation of water: predictions of optimal dopant species from first-principles, J. Phys. Chem. C 121 (2017) 19201-19208.

[48] L.R. Sheppard, S. Hager, J. Holik, R. Liu, S. Macartney, R. Wuhrer, Tantalum segregation in ta-doped $TiO_2$ and the related impact on charge separation during illumination, J. Phys. Chem. C 119 (2015) 392-400.

[49] X. Li, J. Li, G. Zhang, W. Yang, L. Yang, Y. Shen, Y. Sun, F. Dong, Enhanced reactant activation and transformation for efficient photocatalytic acetone degradation on $SnO_2$ via Hf doping, Adv. Sustain. Syst. 5 (2021) 26-29.

[50] M. N. Shaddad, D. Cardenas-Morcoso, M. Garcia-Tecedor, F. Fabregat-Santiago, J. Bisquert, A. M. Al-Mayouf, S. Gimenez, $TiO_2$ Nanotubes for Solar Water Splitting: Vacuum Annealing and Zr doping enhance water oxidation kinetics, ACS Omega 4 (2019) 1609516102.